\documentstyle[aps]{revtex}
\draft
\begin{document}
\title{LOW TEMPERATURE VISCOSITY IN ELONGATED FERROFLUIDS}
\author{T. Alarc\'on, A. P\'erez-Madrid and J. M. Rub\'{\i}}
\address{Departament de F\'{\i}sica Fonamental\\
Facultat de F\'{\i}sica\\
Universitat de Barcelona\\
Diagonal 647, 08028 Barcelona, Spain\\
}

\maketitle

\begin{abstract}
We have studied the relaxation and transport properties of a ferrofluid in an elongational flow. These properties are influenced by the bistable nature of the potential energy. Bistability comes from the irrotational character of the flow together with the symmetry of the dipoles. Additionally, the presence of a constant magnetic field destroys the
symmetry of the potential energy magnetizing the system. We have shown that at a moderate temperature, compared to the
height of the energy barrier, the viscosity decreases with \mbox{respect} to the
value it would have if the potential were stable. This phenomenon is known as
the 'negative viscosity' effect. Thermal motion induces jumps of the
magnetic moment between the two stable states of the system leading to
the aforementioned lowered dissipation effect.
\end{abstract}


\section{Introduction}

The existence of a 'negative' viscosity in dispersions of ferromagnetic particles in a nonpolar solvent is a curious phenomenon which has recently been discovered. It was predicted by Shliomis and Morozov$^{\ref{bib:shliomis}}$ and corroborated experimentally by Bacri $\it{et\; al.}$$^{\ref{bib:bacri}}$. Its interest is based upon the fact that contrary to what one would expect, the viscosity of the dispersion as a whole diminishes when an applied magnetic field oscillates at an optimum frequency. As has been known for a long time, the presence of a constant magnetic field prevents free rotation of the dipoles which originates a resistance on the flow of the fluid which increases dissipation. On the other hand, Bacri $\it{et\; al.}$$^{\ref{bib:bacri}}$ have shown that the increment in the viscosity due to the rotational degrees of freedom is proportional to the difference between the vorticity and the angular velocity of the particles in suspension. Thus, a magnetic field oscillating at a high enough frequency can impart an angular velocity to the particles greater than the vorticity, leading to a 'negative' viscosity. One see that the particles gain kinetic energy at the expense of the oscillating field.
 
This striking phenomena is the motivation for undertaking this investigation. We have found that the effect described is more general than was thought, considering that it can be applied to any ferrofluid that has a frequency different from the inverse of the Brownian relaxation time. 

The key point in our results comes from the fact that we work with axisymmetric particles immersed in a suctioning current. Thus, at a low enough temperature, due to thermal agitation, the particles jump back and forth along the axis of the
flow, which thereby introduces an internal frequency in the system. This internal
frequency plays exactly the same role as the frequency of the alternating magnetic
field in the systems studied by Shliomis and Morozov$^{\ref{bib:shliomis}}$ and Bacri $\it{et\; al.}$$^{\ref{bib:bacri}}$.

There are examples of systems other than ferrofluids that potentially can show this phenomenology, these are suspensions of gravity dipoles$^{\ref{bib:brenner},\ref{bib:brenner2}}$ or suspensions of magnetotactic     bacteria$^{\ref{bib:blakemore}}$. These bacteria, some of them rod shaped, undergo the phenomenon of magnetotaxis. The bacteria contain iron particles which impart a magnetic moment to themselves.
  
The paper is organized as follows; in section II we describe the system and
perform an analysis of the stability of the equilibrium orientations of the dipoles. Section III is
devoted to the derivation of the relaxation equation for the magnetization. In
section IV we introduce fluctuations in our analysis and, by applying
fluctuating hydrodynamics in the space of orientations, we derive the
Langevin equation for the magnetic moment. In section V we compute
the stress tensor and the viscosity tensor. Finally, in section
VI, we discuss our main results.

\section{The Magnetic Rotor}

The system we want to study consists of a dilute colloidal suspension
of ferromagnetic rodlike particles immersed in a nonpolar fluid phase. This suspension flows in an
elongational flow and under the influence of a constant magnetic field $
\vec{H}$ oriented in the direction of the symmetry axis of the flow, which
will be taken parallel to the z-axis. All of the particles
are supposed to have the same magnetic moment

\begin{equation}  \label{eq:a1}
\vec{m}=m_s\hat{R},
\end{equation}

\noindent where $\hat{R}$ is the unit vector along the direction of the axis
of the particle and $m_s$ is the magnetic moment strength. The velocity
field of the flow is given by

\begin{equation}  \label{eq:a2}
\vec{v}=\vec{\vec{\beta}}\cdot\vec{r},
\end{equation}

\noindent where $\vec{\vec{\beta}}$ is the velocity gradient of the flow

\begin{equation}  \label{eq:a3}
\vec{\vec{\beta}}=\beta \left( 
\begin{array}{ccc}
-1 & 0 & 0 \\ 
0 & -1 & 0 \\ 
0 & 0 & 2
\end{array}
\right)
\end{equation}

\noindent with $\beta$ being the rate of elongation.

A suspended particle in the carrier fluid experiences the hydrodynamic
torque given by

\begin{equation}  \label{eq:a4}
\vec{T}^{h} = - \vec{\vec{\xi}}\cdot(\vec{\omega}-\vec{\Omega}),
\end{equation}

\noindent where the friction tensor $\vec{\vec{\xi}}$ is given in terms of its components $\xi_0$ and $\xi_1$

\begin{equation}  \label{eq:a5}
\vec{\vec{\xi}} = \xi_{1} \hat{R}\hat{R} + \xi_{0} (\vec{\vec{1}} - \hat{R}
\hat{R}).
\end{equation}

\noindent Here $\vec{\omega}$, $\vec{\Omega}$ and $\vec{\vec{1}}$ are the
angular velocity of the particle, the drag angular velocity due to the
motion of the fluid and the unit tensor, respectively. Due to the fact that
the only possible relative motion between the ends of the rod is a rigid
rotation one has

\begin{equation}  \label{eq:a6}
\vec{v}_+-\vec{v}_- = \vec{\Omega}\times \hat{R} L,
\end{equation}

\noindent where $\vec{v}_{\pm}$ are the velocities of the end points of
the rod and $L$ is its length. This expression can alternatively be written as

\begin{equation}  \label{eq:a7}
\vec{v}_+-\vec{v}_- = \vec{\vec{\beta}}\cdot\hat{R} L.
\end{equation}

\noindent Thus, through a comparison of eqs. (\ref{eq:a6}) and (\ref{eq:a7})
we achieve

\begin{equation}  \label{eq:a8}
\vec{\Omega}
 = \vec{{\cal {R}}}(1/2 \vec{\vec{\beta}} : \hat{R}\hat{R}),
\end{equation}

\noindent with $\vec{{\cal {R}}} \equiv \hat{R}\times\frac{\partial}{
\partial \hat{R}}$ being the rotational operator.

In addition to eq. (\ref{eq:a4}) the dipole is acted upon by a magnetic
torque

\begin{equation}  \label{eq:a9}
\vec{T}^m = \vec{m}\times\vec{H}.
\end{equation}

\noindent Therefore, by taking into account the angular momentum conservation, in view of
eqs. (\ref{eq:a4}), (\ref{eq:a5}), (\ref{eq:a8}), and (\ref{eq:a9}) we can write in the high friction limit

\begin{equation}  \label{eq:a10}
\vec{\omega} = \frac{1}{\xi_0}(-\vec{{\cal {R}}}U),
\end{equation}

\noindent where the potential U, defined through

\begin{equation}  \label{eq:a11}
U = -\vec{m}\cdot\vec{H} - \frac{1}{2}\xi_0\vec{\vec{\beta}}: \hat{R}\hat{R};
\end{equation}

\noindent accounts for the different mechanisms for the rotation of the
particle: the magnetic field and the external flow. In polar spherical
coordinates this energy can be rewritten as

\begin{equation}  \label{eq:a12}
U=-m_sH\cos{\theta}+\frac{\xi_0\beta}{2}(1-3\cos^2{\theta}),
\end{equation}

\noindent $\theta$ being the angle between the axis of the particle and the
z-axis. 

The equilibrium states of the system can be identified through the condition $
\frac{dU}{d\theta}=0$, i.e.

\begin{equation}  \label{eq:a13}
m_{s}H\sin\theta+3\xi_0\beta\sin\theta\cos\theta=0
\end{equation}

\noindent whose solutions, $\theta_-$, $\theta_+$ and $\theta_0$, are

\begin{equation}  \label{eq:a14}
\theta_{-}=0,
\end{equation}

\begin{equation}  \label{eq:a15}
\theta_{+}=\pi,
\end{equation}

\begin{equation}  \label{eq:a16}
\theta_0= \arccos \left (-\frac{m_sH}{3\xi_0\beta}\right ).
\end{equation}

\noindent The stability of these orientations follows from the second derivative of the potential computed in each of the solutions 

\begin{equation}  \label{eq:a18}
\frac{d^{2}U}{d\theta^2}\vert_{\theta_-}=m_{s}H+3\xi_0\beta=m_{s}H(1+
\frac{H_{c}}{H}),
\end{equation}

\begin{equation}  \label{eq:a19}
\frac{d^{2}U}{d\theta^2}\vert_{\theta_+}=-m_{s}H+3\xi_0\beta=m_{s}H(
\frac{H_{c}}{H}-1),
\end{equation}

\begin{equation}  \label{eq:a20}
\frac{d^{2}U}{d\theta^2}\vert_{\theta_0}=\frac{(m_{s}H)^2}{3\xi_0\beta}
-3\xi_0\beta.
\end{equation}

\noindent Since $m_s$, $H$, $\xi_0$ and $\beta$ are always positive
quantities, after examining eqs. (\ref{eq:a18})-(\ref{eq:a20}) we 
conclude that provided

\begin{equation}  \label{eq:a21}
H\leq H_c,
\end{equation}

\noindent $H_c \equiv 3\xi_0\beta/m_s$ being a critical field, $\theta_-$ and $
\theta_+$ are both stable, whereas $\theta_0$ is unstable. We then conclude that under these conditions the potential of the magnetic rotor is bistable. For magnetic fields larger than $H_c$ the system becomes stable.

Equivalently, we could have considered the presence of a critical value for
the elongational rate, $\beta_{c}=\frac{Hm_{s}}{3\xi_0}$, such that for a
fixed value of the magnetic field the system is bistable whenever the actual
value of the elongational rate $\beta$ overcomes $\beta_c$. In the rest of the paper we will assume that $H/H_c < 1$, i. e. we will
remain in the bistable region, or in the range of large elongational rate.

To conclude this section we will give some estimates of the critical field corresponding to situations of experimental accesibility. For particles of magnetite having a volume $V_p = 5\times 10^{-19} cm^3$ and an aspect ratio $\epsilon = 0.1$ one has $m_s = 2.4\times 10^{-16} G\times cm^3$, and consequently $H_c = 58.3 Oe$. On the other hand, for particles of cobalt with volume $V_p = 2.7\times 10^{-19} cm^3$ and $\epsilon = 0.1$ we obtain $m_s = 3.8\times 10^{-16} G\times cm^3$, and $H_c = 17.1 Oe$. In both cases we have assumed $\beta = 10^3 s^{-1}$. At smaller elongational rate, $\beta = 10 s^{-1}$, one has $H_c = 5.8 Oe$ for magnetite and $H_c = 1.7 Oe$ for cobalt.

\section{Relaxation equation}

The relaxation of the magnetic rods can be interpreted as a diffusion process through
a potential barrier in orientation space. In this scenario we can apply the
formalism of non-equilibrium thermodynamics$^{\ref{bib:reacciones}}$. Let $
\rho(\hat{R},t)$ be the angular distribution function that may be viewed as a
density in the space of orientations. Thus, we can define the chemical
potential as

\begin{equation}  \label{eq:b1}
\mu=K_{B}T\ln{\rho}+U.
\end{equation}

\noindent The diffusion equation for $\rho$ is written in the form

\begin{equation}  \label{eq:b2}
\frac{\partial\rho}{\partial t}=\frac{D}{\sin\theta}\frac{\partial}{
\partial\theta}\{\sin\theta[\frac{\rho}{K_{B}T}\frac{\partial U}{
\partial\theta}+\frac{\partial\rho}{\partial\theta}]\},
\end{equation}

\noindent with $D = k_B T/\xi_0$ and where the axial symmetry of the
potential (\ref{eq:a12}) has been taken into account. Equation (\ref{eq:b2})
can be rewritten as a continuity equation

\begin{equation}  \label{eq:b3}
\frac{\partial\rho}{\partial t}=-\frac{1}{\sin\theta}\frac{\partial}{
\partial \theta}(J_{\theta}\sin\theta ),
\end{equation}

\noindent which defines the diffusion current

\begin{equation}  \label{eq:b4}
J_{\theta}=-De^{(-U/K_{B}T)}\frac{\partial}{\partial\theta}e^{\mu /K_{B}T}.
\end{equation}

If the height of the potential barrier is large enough as compared to
thermal energy $K_{B}T$, we can suppose that equilibrium is reached
independently on each side of the barrier. Thus, the chemical potential can
be inferred as being

\begin{equation}  \label{eq:b5}
\mu(\hat{R},t)=\mu(\theta_{-})\Theta (\theta_{0}-\theta)+\mu(\theta_{+})\Theta (\theta-
\theta_{0}),
\end{equation}

\noindent where $\Theta(\theta)$ is the unit step function. The system is
allowed to obtain the global equilibrium due to the presence of a
quasi-stationary current $J(t)$, assumed uniform.

\begin{equation}  \label{eq:b6}
J_{\theta}\sin\theta =J(t)\{\Theta(\theta-\theta_{-})+\Theta(\theta-\theta_{+})\}.
\end{equation}

\noindent This adiabatic hypothesis is one of the essential points in the
present development. By using (\ref{eq:b1}) and (\ref{eq:b5}), one has

\begin{equation}  \label{eq:b7}
\rho(\hat{R},t)=\rho_{-}(t)e^{[-(U-U_{-})/K_{B}T)]}\Theta(\theta_{0}-\theta)+
\rho_{+}(t)e^{[-(U-U_{+})/K_{B}T)]}\Theta(\theta-\theta_{0}),
\end{equation}

\noindent where $\rho_{\pm}\equiv\rho( \theta_{\pm},t)$, and $U_{\pm}\equiv
U(\theta_{\pm})$, are the densities and potential energies in the two stable
states. In order to obtain an expression for $J(t)$, we substitute (\ref{eq:b4})
into (\ref{eq:b6}) to arrive at

\begin{equation}  \label{eq:b8}
J(t)\frac{e^{U/K_{B}T}}{\sin\theta}\{\Theta(\theta-\theta_{-})+\Theta(\theta-
\theta_{+} )\}=-D\frac{\partial}{\partial\theta}e^{{\mu}/K_{B}T}.
\end{equation}

\noindent Integrating now over $\theta$ and taking into account that, due to
the height of the barrier, the main contribution to these integrals is
around the maximum of the potential $\theta_0$, one has the law of mass      action$^{\ref{bib:mazur}}$

\begin{equation}  \label{eq:b9}
J(t)=K_{B}{\it {l}(1-e^{A/K_{B}T}),}
\end{equation}

\noindent which is a nonlinear phenomenological relationship between the
quasi-stationary current $J(t)$ and the affinity $A\equiv\mu_{+}-\mu_{-}$, where $\mu_{\pm}\equiv\mu (\theta_{\pm})$. The phenomenological coefficient ${\it {l}}$ is given by

\begin{equation}  \label{eq:b10}
{\it {l}=\frac{D\sin\theta_0}{K_B}\rho_{-}(\frac{\vert
U_{0}^{^{\prime\prime}}\vert}{2\pi K_{B}T})^{1/2}e^{(U_{0}-U_{-})/K_{B}T},}
\end{equation}

\noindent where $U_{0}^{^{\prime\prime}}\equiv\frac{d^{2}U}{d\theta^2}
\vert_{\theta_0}$ and $U_{0}\equiv U(\theta_{0})$.

We are now prepared to proceed to the deduction of the relaxation
equations. To this end, we define the following populations:

\begin{equation}  \label{eq:b12}
N_{+}\equiv\int_{\phi=0}^{2\pi}\int_{\theta=\theta_0}^{\pi}\rho(\hat{R},t)d
\hat{R}=\int_{\theta_0}^ {\pi}2\pi\sin \theta\rho (\hat{R} ,t)d\theta,
\end{equation}

\noindent and

\begin{equation}  \label{eq:b13}
N_{-}\equiv\int_{\phi=0}^{2\pi}\int_{\theta=0}^{\theta_0}\rho(\hat{R},t)d
\hat{R}=\int_{0}^{\theta_0}2\pi\sin \theta\rho (\hat{R} ,t)d\theta,
\end{equation}

\noindent related by the normalization condition 
\begin{equation}  \label{eq:l1}
N = N_+ + N_{-}.
\end{equation}

\noindent The kinetic equations for $N_+$ and $N_-$ follow after differentiating the 
eqs. (\ref{eq:b12}) and (\ref{eq:b13}) and employing eq. (\ref{eq:b3}).
Thus, we have

\begin{equation}  \label{eq:b14}
\frac{dN_+}{dt}=-\frac{dN_-}{dt}=2\pi J(t),
\end{equation}

\noindent where, consistently with the adiabatic approximation

\begin{equation}  \label{eq:b15}
N_{-}=\frac{2\pi K_{B}T}{U_{-}^{^{\prime\prime}}}\rho_-,
\end{equation}

\begin{equation}  \label{eq:b16}
N_{+}=\frac{2\pi K_{B}T}{U_{+}^{^{\prime\prime}}}\rho_+,
\end{equation}

\noindent in which $U_{\pm}^{^{\prime\prime}}\equiv\frac{d^2U}{d\theta^2}
\vert_{\theta_{\pm}}$. Using equations (\ref{eq:b9}) and (\ref{eq:b14})-(\ref
{eq:b16}) we can derive the rate equations for the two populations

\begin{equation}  \label{eq:b17}
\frac{dN_+}{dt}=-\frac{dN_-}{dt}=K_{+-}N_{-}-K_{-+}N_+,
\end{equation}

\noindent where the rate constants are given by

\begin{equation}  \label{eq:b18}
K_{+-}=\frac{D\sin\theta_0}{2\pi K_{B}T}(\frac{\vert
U_{0}^{^{\prime\prime}}\vert}{2\pi K_{B}T})^{1/2}
U_{-}e^{-(U_{0}-U_{-})/K_BT},
\end{equation}

\noindent and

\begin{equation}  \label{eq:b19}
K_{-+}=\frac{D\sin\theta_0}{2\pi K_BT}(\frac{\vert
U_{0}^{^{\prime\prime}}\vert}{2\pi K_{B}T})^{1/2}U_+^{^{\prime
\prime}}e^{-(U_0-U_+)/K_BT}.
\end{equation}

We can now define the magnetic moment parallel to the direction of the
applied magnetic field as

\begin{equation}  \label{eq:b20}
m\equiv m_{s}\frac{N_{+}-N_-}{N}.
\end{equation}

\noindent By differentiating equation (\ref{eq:b20}) and employing equation 
(\ref{eq:b17}) we obtain the relaxation equation for ${\it {m}}$

\begin{equation}  \label{eq:b21}
\frac{dm}{dt}=-\frac{1}{\tau}m+\frac{1}{\alpha}m_s,
\end{equation}

\noindent where

\begin{equation}  \label{eq:b22}
\tau=\frac{1}{K_{+-}+K_{-+}}
\end{equation}

\noindent is the relaxation time, and

\begin{equation}  \label{eq:b24}
\alpha=\frac{1}{K_{+-}-K_{-+}}
\end{equation}

\noindent accounts for the asymmetry of the potential. An equation similar to (\ref{eq:b21}) was postulated by Shliomis in the context of              ferrohydrodynamics$^{\ref{bib:shliomis2}}$. Our approach to deduce eq. (\ref{eq:b21}) has been based upon mesoscopic arguments and as we will see in the next section it provides the natural way of introducing fluctuations in the scheme.

It is useful 
for our purposes to introduce the nondimensional
 variables $x\equiv\frac{H_c
}{H}$ and $\mu\equiv\frac{m_sH}{K_BT}$, in terms of which the relaxation
time is written as
\pagebreak

$$
\tau=\frac{(2\pi)^{3/2}}{D}(x\mu)^{(-1/2)}(1-\frac{1}{x^2})^{-1}\exp\{ \frac{\mu}{2}(\frac{1}{x}+$$
\begin{equation}\label{eq:b25}
\frac{x}{3})\}[\mu(x-1)\exp
\{\mu(1-\frac{x}{3})\}+\mu(x+1)\exp\{-\mu(1+\frac{x}{3})\}]^{-1},
\end{equation}

\noindent where $x$ must be greater than the unity. The inverse of $\tau$
gives us a characteristic frequency proper to the system, the jump frequency
between the two stable states of the potential energy. This frequency is
given by

\begin{equation}  \label{eq:b26}
\omega=D(\frac{x\mu}{2\pi})^{3/2}(1-\frac{1}{x^2})e^{-\frac{x\mu}{2}}e^
{-\mu/x}[e^{\mu}+e^{-\mu}+\frac{1}{x}(e^{-\mu}-e^{\mu})].
\end{equation}

In Figure 1 we can observe the behavior of $\omega$ when one varies the
elongational rate. The way in which it depends on the elongational rate and
the existence of two time scales in our problem, Brownian $D^{-1}$ and $
\tau$, are the key points in understanding the dynamical mechanism which
leads to the results we will obtain in section 5.

\section{Dynamic of the fluctuations of the magnetic moment}

Thermal motion of the dipoles inside the carrier fluid produces fluctuations
in the population of the minima of the potential. This fact manifests itself
on a mesoscopic level through fluctuations of the magnetic moment. These can
be taken into account by adding a random current$^{\ref{bib:reacciones}}$ to equation (\ref{eq:b9}),

\begin{equation}  \label{eq:c1}
J(t)=K_B{\it {l}(1-e^{A/K_BT})+\frac{\sqrt{\bar{D}}}{m_s}\xi(t),}
\end{equation}

\noindent where $\bar{D}$ is given by

\begin{equation}  \label{eq:c2}
\bar{D}=2m_{s}^{2}k_B\bar{{\it {l}}},
\end{equation}

\noindent $\bar{l}$ is the equilibrium value of the phenomenological
coefficient ${\it {l}}$, which according to (\ref{eq:b10}) should be
proportional to $\rho_-$ computed at equilibrium, $\rho_{-}^{eq}$. By
applying the detailed balance principle to eq. (\ref{eq:b17}) and using the
normalization relation eq. (\ref{eq:l1}) one can easily achieve

\begin{equation}  \label{eq:l2}
\rho_{-}^{eq} =\left ( \frac{N}{2\pi k_BT}\right )\;
\frac{U_{-}^{^{\prime\prime}}U_{+}^{^{
\prime\prime}}}{U_{+}^{^{\prime\prime}} 
+ U_{-}^{^{\prime\prime}}\exp\{\frac{
U_- - U_+}{k_BT}\}}.
\end{equation}

\noindent Additionally, in eq. (\ref{eq:c1}) $\xi (t)$ is a Gaussian white
noise stochastic process of zero mean and correlation function.

\begin{equation}  \label{eq:c3}
\langle\xi(t)\xi (t^{\prime})\rangle=\delta (t-t^{\prime}).
\end{equation}

From equations (\ref{eq:b14}), (\ref{eq:b20}) and (\ref{eq:c1}) we obtain
the Langevin equation for m

\begin{equation}  \label{eq:c4}
\frac{dm}{dt}=-\frac{1}{\tau}m+\frac{1}{\alpha}m_s
+\sqrt{\bar{D}}\xi(t).
\end{equation}

\noindent Notice that coefficient $\bar{D}$ is the input noise strength
corresponding to the stochastic process $m$. This coefficient can by
explicitly written as

\begin{equation}\label{eq:c8}
\bar{D}=m_{s}^2D\sin\theta_0(\frac{\vert U''_{0}\vert}{2\pi K_BT})^{1/2} 
(1+\frac{\tau}{\alpha})\frac{U''_+}{2\pi K_BT}exp\{-\frac{(U_0-U_+)}{K_BT}\}
\end{equation}

\noindent and in terms of the nondimensional parameters $x$ and $\mu$

\begin{equation}\label{eq:c9}
\bar{D}=\frac{m_{s}^2D}{(2\pi)^{3/2}}(x\mu)^{1/2}(1-\frac{1}{x^2})\mu(x-1)\frac{2(1+x)e^{-\mu}}{(1+x)e^{-\mu}+(x-1)e^{\mu}}                  
\exp\{-\frac{\mu}{2}(\frac{1}{x}+\frac{x}{3})\}exp\{\mu(1-\frac{x}{3})\}.
\end{equation}

\noindent Finally, in reference to eq. (\ref{eq:c4}), by performing the change

\begin{equation}  \label{eq:c5}
\tilde{m}=m-\frac{\tau}{\alpha}m_s,
\end{equation}

\noindent this equation becomes

\begin{equation}  \label{eq:c6}
\frac{d\tilde{m}}{dt}=-\frac{1}{\tau}\tilde{m}
+\sqrt{\bar{D}}\xi (t),
\end{equation}

\noindent i.e., $\tilde{m}$ is an Ornstein-Uhlenbeck process$^{\ref{bib:Gardiner}}$. As is well known, the stationary distribution for such a
process is

\begin{equation}  \label{eq:c7}
p(\tilde{m})=\frac{1}{\sqrt{\pi\tau\bar{D}}}
\exp\{-\frac{\tilde{m}^2}{\tau
\bar{D}}\},
\end{equation}

\noindent which will be used in the next section to compute the viscosity.

\section{The Viscosity Tensor}

In order to calculate the viscosity tensor, we first have to compute  the stress
tensor. This quantity has two contributions, one that comes from the solvent, and the other due to the presence of the particles. The latter is written$^{\ref{bib:Doi}}$ as

\begin{equation}  \label{eq:d1}
\vec{\vec{\sigma}}=nK_BT(3\langle\hat{R}\hat{R}\rangle-
\vec{\vec{1}})+ n
\xi_1\vec{\vec{\beta}}
:\langle{\hat{R}\hat{R}\hat{R}
\hat{R}}\rangle-nK_BT\mu\langle\hat{R}\hat{H}
\cdot(\vec{\vec{1}}-\hat{R}\hat{
R})\rangle,
\end{equation}

\noindent where $n$ is the concentration of suspended particles and $L$ is
their longitude. Thus, eq. (\ref{eq:d1}) give us first order contributions
to the viscosity tensor. The moments in eq. (\ref{eq:d1}) will be calculated
by using the stationary distribution (\ref{eq:c7}). It is important to keep
in mind that although we are using a stationary distribution, the dynamical
effects are taken into account through their dependence on the relaxation
time $\tau$.

\noindent We will illustrate the behavior of the viscosity by explicitly computing  the parallel viscosity defined through 

\begin{equation}\label{eq:d3}
\eta_{\parallel\parallel} = \frac{\sigma_{\parallel\parallel}}{\beta}
\end{equation}

\noindent with $\sigma_{\parallel\parallel} = \hat{H}\cdot\vec{\vec{\sigma}}
\cdot\hat{H}$, that from eq. (\ref{eq:d1}) is written as

\begin{equation}  \label{eq:d4}
\sigma_{\parallel\parallel}=
nK_BT(3\langle \hat{R}_{\parallel}\hat{R}_{\parallel}\rangle
-1)+ n\xi_1
\beta(\langle \hat{R}_{\parallel}\hat{R}_{\parallel}
\rangle-\langle \hat{R}_{\parallel}\hat{R}_{\parallel}
\hat{R}_{\parallel}\hat{R}_{\parallel}\rangle
) -nK_BT\mu(\langle \hat{R}_{\parallel}\rangle-\langle
\hat{R}_{\parallel}\hat{R}_{\parallel}\hat{R}_{\parallel}\rangle).
\end{equation}

\noindent Here we have taken into account that $\vec{\vec{\beta}}:\hat{R}\hat{R}=\beta(1-3\hat{R}
_{\parallel}^2)$, and $\hat{R}_{\parallel\parallel}= \hat{R}\cdot\hat{H}\hat{H}$. In the appendix we summarize the result of the computation of all the
moments that appear in the expression for 
 $\sigma_{\parallel\parallel}$. Making use of these results, we find that

$$\sigma_{\parallel\parallel}=nK_BT[\frac{\tau\bar{D}}{2m_{s}^2}+
(\frac{\tau}{\alpha})^2-1]+
n\xi_1\beta[(\frac{\tau}{\alpha})^2+
\frac{\tau\bar{D}}{2m_{s}^2}- 
9(\frac{\tau\bar{D}}
{2m_{s}^2})^2-$$\\
\begin{equation}\label{eq:d5}
18\frac{\tau}{\alpha}\frac{\tau\bar{D}}{2m_{s}^2}-
3(\frac{\tau}{\alpha})^4] - nK_BT\mu[\frac{\tau}{\alpha}-
(\frac{\tau}{\alpha})^3-
3\frac{\tau}{\alpha}\frac{\tau\bar{D}}{2m_{s}^2}].
\end{equation}

\noindent Thus, we finally obtain
\pagebreak

$$\frac{\eta_{\parallel\parallel}}{3n\xi_0}=
\frac{1}{x\mu}[(1+\mu)\frac{\tau}{\alpha}+\frac{\tau\bar{D}}{2m_{s}^2}+
3\mu\frac{\tau}{\alpha}\frac{\tau\bar{D}}{2m_{s}^2}+
\mu(\frac{\tau}{\alpha})^3-1] +$$\\ 
\begin{equation}\label{eq:d6}
\frac{1}{3}\frac{\xi_1}{\xi_0}[\frac{\tau\bar{D}}{2m_{s}^2}-18\frac{\tau}{\alpha}\frac{\tau\bar{D}}{2m_{s}^2}-9(\frac{\tau\bar{D}}{2m_{s}^2})^2+(\frac{\tau}{\alpha})^2-3(\frac{\tau}{\alpha})^4].
\end{equation}

\noindent In figures 2 and 3 we have plotted the nondimensional
quantity $\eta_{par}\equiv\frac{\eta_{\parallel\parallel}}{3n\xi_0}$ in terms of $x$ and $
\omega$, respectively, for particles with an aspect ratio $\epsilon=0.1$,
for which one has $\frac{1}{3}\frac{\xi_1}{\xi_0} = 0.1805$$^{\ref{bib:viscositat}}$. 

One can observe that this viscosity becomes negative for
large shear rates and saturates at a positive value. Similar results have been obtained by Shliomis and Morozov$^{\ref{bib:shliomis}}$ and                                       Bacri ${\it {et\;al.}}$$^{\ref{bib:bacri}}$ with
ferrofluids under an oscillating magnetic field. Negative viscosities in
these cases were due to the  existence of two
characteristic time scales in the system that enter in competition. One of them is related to the frequency of the alternating magnetic field and the other is related to the vorticity of the fluid. In our case, the dipoles relax to the equilibrium orientations in a time scale $D^{-1}$, which is shorter than the relaxation time $\tau$ associated with the diffusion through the potential barrier. The rods are then constantly jumping with a frequency $\omega =\tau^{-1}$  acquiring a net angular velocity different from zero, whose expression is given in the appendix. The existence of this angular velocity then shows the conversion of thermal energy into kinetic energy for rotation. This explains why, in our case, the viscosity diminishes.

In looking at fig. 3 one first observes a histeresys cycle described by $\eta_{par}$ when
$\omega$ varies. During the first stage, when the actual value of $\omega
$ is between zero and its maximum value, $\eta_{par}$,
decreases from a positive value to a minimum negative value, as occurs in
the case analyzed by Shliomis and Morozov$^{\ref{bib:shliomis}}$. These results are in good
agreement with those found by them. The difference between our approach
and the one by Shliomis and Morozov$^{\ref{bib:shliomis}}$ is that the externally imposed
frequency can be as large as you want. In our case the frequency of
relaxation, the result of an internal dynamical mechanism, does not grow
freely. In the second stage; $\omega$ is a decreasing function of the
shear rate. This fact is responsible for the particular behavior of $
\eta_{par}$ observed. Note that, as can be seen in Figure 2, $
\eta_{par}$ does not saturate to a negative value when $
\beta\rightarrow\infty$. It tends very slowly to a positive value, as
slowly as $\omega$ tends to zero. Our results also agree with the analysis
of the phenomenon carried out by Rosensweig$^{\ref{bib:rosensweig}}$ as well.

\section{Conclusions}

In this paper we have shown that in a ferrofluid made from rodlike particles under a constant magnetic field and in a elongational flow, rotations of the particles lead to
non-monotonous behavior of the viscosity. This fact allows us to generalize the phenomenon discovered by Bacri $\it{et al.}$$^{\ref{bib:bacri}}$.

As a consequence of the orientating effect of the flow, the particles rotate
with a drag angular velocity $\vec{\Omega}$ that eliminates the degeneracy
of the direction of their actual angular velocity in a volume element of the
ferrofluid (see eq. (\ref{eq:a10})). Thus, all the contributions to
dissipation in a volume element add up constructively.

Moreover, because the flow is elongational, we can write a potential energy
of orientation related to it. This potential is bistable and the addition of a magnetic
field breaks up its symmetry, thus magnetizing the system. On the other hand, thermal motion causes jumps
between the two stable states of the potential with a certain frequency. Consequently, the particles acquire a kinetic energy for rotation at the expense of thermal energy. This
fact eliminates the rigidity that the system would have if no motion were
present, i.e. if the potential time independent were stable. In fact,
this is the way that we interpret the 'negative viscosity' effect predicted
for our system.

We have derived the kinetic equations for the population of the minima of the
potential and from it the relaxation equation for the magnetic moment.  Likewise,
by assuming fluctuations of the density of states in the orientation space, we
have formulated the Langevin equation for the magnetic moment. This equation
describes a Ornstein-Uhlembeck process whose moments are well known. The
computation of the first four moments allows us to  derive the viscosity.

It should be emphasized that the viscosity is computed in the limit of high energy barrier, i.e. for strong rate of elongation. This is the opposite situation to the case studied in the two previous papers$^{\ref{bib:viscositat},\ref{bib:dumbbell}}$ where we covered the weak flow regime, showing an increase of the viscosity. Analogously, modifying the magnetic field without altering the rate of elongation we achieve the same effect, that is, to vary the viscosity. Thus, a possible application of our results is in adaptive dampers. 

Through this behavior of the viscosity, emerge the macroscopic consecuences of the dynamical bifurcation with exchange of stability that our system experiences.

Likewise, the effect that we are studying is only possible in dilute solutions,where each rod can rotate freely without interference by others. Of course in neglecting hydrodynamic interactions among the particles we assume a extremely dilute ferrofluid. Anyway, this is the first step in our study of nonlinear or hysteresis effects in the rheology of ferrofluids suspensions. 

The following step will be to assume higher concentrations. Nonetheless, to include hydrodynamic and exclude volume interactions in the dynamics of an assembly of rod particles is difficult. So that, we will model the elongated dipoles by means of rigid dumbbells, such models are well studied in the field of polymeric liquids.

In order to increase the modulation effect of the magnetic field or of the elongation rate on the viscosity, another possibility we are thinking of consists in adding nonmagnetic spheres at high concentration to the ferrofluid.

\acknowledgments

This work has been supported by DGICYT of the Spanish Government under grant
PB95-0881, and also by the INCO-COPERNICUS program of the European Commission under contract IC15-CT96-0719. One of us (T. Alarc\'on) wishes to thank to DGICYT of the Spanish Government for financial support.

\appendix
\section{}

To begin with, in this appendix we give the moments which must be calculated in order to
compute the stress tensor.

\begin{equation}  \label{eq:e1}
\langle \hat{R}_{\parallel}\rangle=\frac{\tau}{\alpha}
\end{equation}

\begin{equation}  \label{eq:l3}
\langle \hat{R}_{\parallel}\hat{R}_{\parallel}\rangle= \frac{\langle\tilde{m}^2\rangle}{
m_{s}^2}+(\frac{\tau}{\alpha})^2= \frac{\tau\bar{D}}{2m_{s}^2}+(\frac{\tau}{
\alpha})^2
\end{equation}

\begin{equation}  \label{eq:e2}
\langle \hat{R}_{\parallel}\hat{R}_{\parallel}\hat{R}_{\parallel}\rangle=3\frac{\tau\bar{D}}{
2m_{s}^2}\frac{\tau}{\alpha}+(\frac{\tau}{\alpha})^3
\end{equation}

\begin{equation}  \label{eq:e3}
\langle \hat{R}_{\parallel}\hat{R}_{\parallel}\hat{R}_{\parallel}\hat{R}_{\parallel}\rangle=3(\frac{
\tau\bar{D}}{2m_{s}^2})^2+6(\frac{\tau}{\alpha})^2\frac{\tau\bar{D}}{2m_{s}^2
}+(\frac{\tau}{\alpha})^4\; .
\end{equation}

Finally, according to eqs. (\ref{eq:a8}), (\ref{eq:a10}) we can write

\begin{equation}\label{eq:l5}
\vec{\omega}-\vec{\Omega} = \frac{1}{\xi_0} \vec{m}\times\vec{H}.
\end{equation}

\noindent Now we define the root mean square angular velocity $\omega_{rm} \equiv \sqrt{\langle(\vec{\omega}-\vec{\Omega})^2\rangle}$, that in view of eq. (\ref{eq:l5}) is given by

\begin{equation}\label{eq:l6}
\omega_{rm} = \mu D \left\{ 1 - \frac{\tau\overline{D}}{2m_s^2} - \left (\frac{\tau}{\alpha}\right )^2\right \}^{1/2}.
\end{equation}

\begin{center}
{\bf BIBLIOGRAPHY}
\end{center}

\begin{enumerate}
\item \label{bib:shliomis}  M. I. Shliomis, I. Morozov, Phys. Fluids               {\bf 6}, 2855 (1994).

\item \label{bib:bacri}  J. C. Bacri, R. Perzynski, M.I. Shliomis, G.              I. Burde, Phys. Rev. Lett. {\bf 75}, 2128 (1995).

\item \label{bib:brenner}  H. Brenner, Int. J. Eng. Sci. {\bf 22}, 645             (1984).

\item \label{bib:brenner2}  H. Brenner, J. Colloid. Interface Sci. {\bf             32}, 141 (1970).

\item \label{bib:blakemore} R. Blakemore, Science {\bf 190}, 377                   (1975). R. Blakemore and R.B. Frankel, Scientific American,                   December 1981, page 58.

\item \label{bib:reacciones}  I. Pagonabarraga, A. P\'erez-Madrid, J.M.
            Rub\'{\i},Physica A {\bf 237}, 205 (1997).

\item \label{bib:mazur} S.R. de Groot and P. Mazur, "Non-Equilibrium               Thermodynamics" (Dover, New York, 1984).

\item \label{bib:shliomis2}  M.I. Shliomish, Sov. Phys. JETP {\bf 34},             1291, (1972).

\item \label{bib:Gardiner}  C. W. Gardiner, "Handbook of Stochastic                Methods",(Springer-Verlag, Berlin, 1990).

\item \label{bib:Doi}  M. Doi, S. F. Edwards, "The Theory of Polymer               Dynamics." (Oxford Univ. Press, New York, 1989).

\item \label{bib:viscositat}  C. Salue$\tilde{n}$a, A. P\'erez-Madrid,              J.M. Rub\'{\i}, J. Colloid Interface Sci. {\bf 164}, 269 (1994).

\item \label{bib:rosensweig}  R. E. Rosensweig, Science {\bf 271}, 614              (1996).

\item \label{bib:dumbbell}  C. Salue$\tilde{n}a$, A. P\'erez-Madrid                and J. M. Rub\'{\i}, J. Chem. Phys. {\bf 96}, 6950 (1992).

\end{enumerate}

\newpage

\vspace{3cm}
\large

\begin{center}
FIGURE CAPTIONS
\end{center}
\begin{itemize}

\item{Figure 1}.- Nondimensional frequency versus the scaled elongational rate $x$, for $\mu = 1$.

\item{Figure 2}.- Normalized viscosity versus the scaled elongational rate $x$, for $\mu = 1$.

\item{Figure 3}.- Normalized viscosity versus the nondimensional jump frequency for $\mu = 1$.\\

\end{itemize}

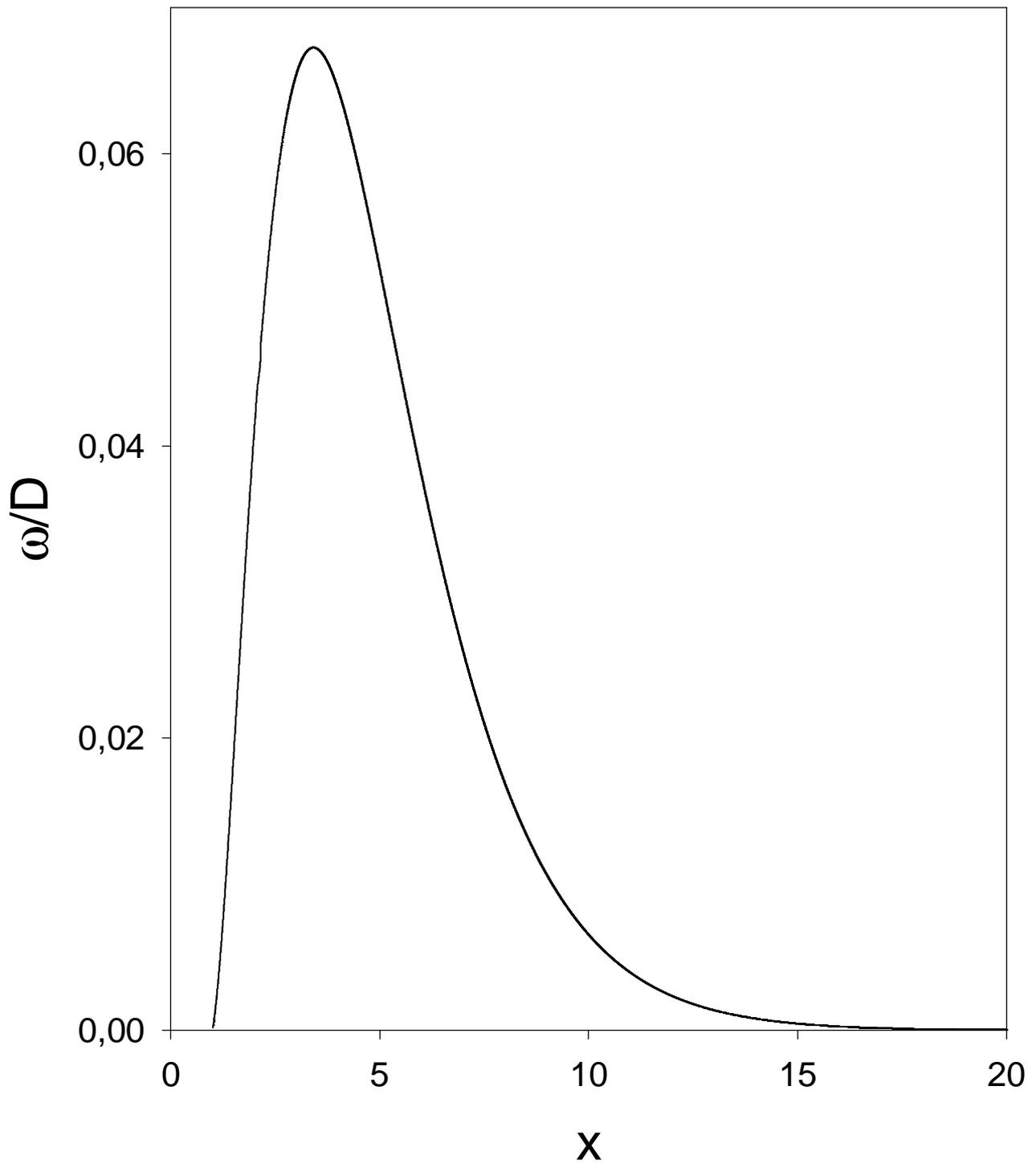
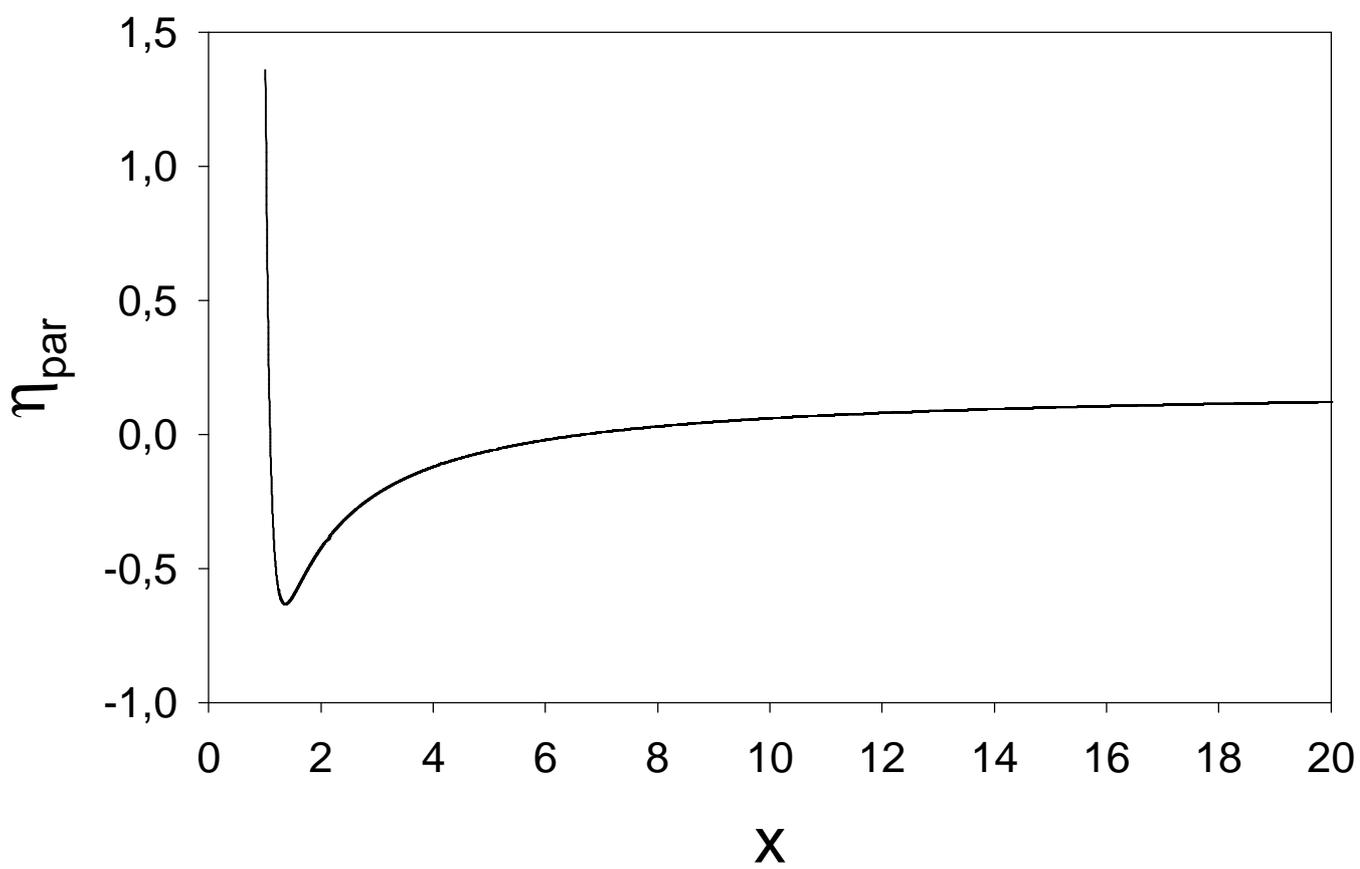
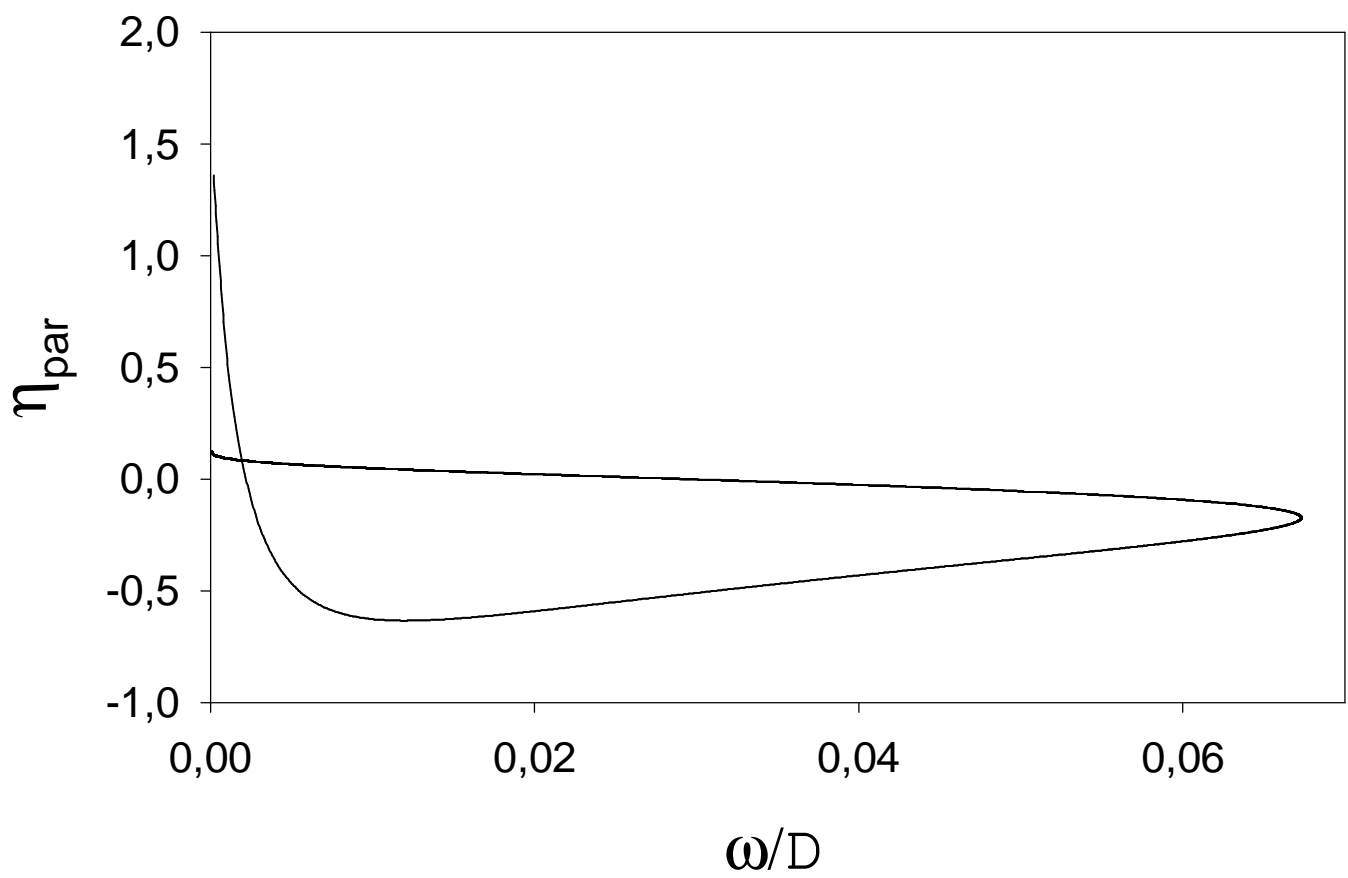
\end{document}